\documentstyle[prb,aps,epsfig]{revtex}

\begin{document}

\newcommand{\kag}{kagom\'{e} }

\newcommand{\um}{\mu m}

\draft

\tighten

\twocolumn[\hsize\textwidth\columnwidth\hsize\csname @twocolumnfalse\endcsname

\title{Comparison of Phase Boundaries between Kagom\'{e} and Honeycomb Superconducting Wire Networks}

\author{Yi Xiao, David A. Huse, Paul M. Chaikin}
\address{Department of Physics, Princeton University,Princeton, NJ 08544}
\author{Mark J. Higgins, Shobo Bhattacharya}
\address{NEC Research Institute, Princeton, NJ 08540}
\author{David Spencer}
\address{Cornell Nanofabrication Facility, Cornell University, Ithaca, NY 14853}
\date{\today}
\maketitle

\begin{abstract}

We measure resistively the mean-field superconducting-normal
 phase boundaries of both \kag and honeycomb wire networks 
immersed in a transverse magnetic field. In addition to their agreement with 
theory about the overall shapes of  phase diagrams, 
they show striking one-to-one correspondence between the cusps in the honeycomb 
and \kag phase boundaries.  
This correspondence is due to their geometric arrangements and agrees with Lin 
and Nori's recent calculation.  We also find that for the frustrated honeycomb
network at $f=1/2$, the current patterns in the superconducting phase 
differ between the low-temperature London regime and the higher-temperature
Ginzburg-Landau regime near $T_c$.
\end{abstract}

\pacs{PACS numbers: 74.80.-g}

\vskip 2pc]

\narrowtext

The complex and interesting properties of superconducting networks
of a variety of geometries in a magnetic field have been extensively  
studied in recent years \cite{1,2,3,4,6,7,8,9,10,11}.
Their properties, as shown in the rich structure of the  
superconducting-normal phase diagram, are found to be very sensitive 
to the topology, and particularly to the connectivity of the structure.
Dips or cusps in the resistively measured transition temperature, 
as a function of the external magnetic field, are indications of the lock-in of a favorable
flux arrangement in the structure. Our experimental results on the
comparison of honeycomb
and \kag networks demonstrate the interesting effects of geometric structure on the
phase boundaries.

The samples we used are aluminum networks fabricated at Cornell Nanofabrication Facility 
with electron-beam lithography. The overall size 
of the samples are $0.8 \times 0.8 mm$. The lattice constant is $1 \um$, with a wire 
width of $0.2 \um$, and a thickness of $50nm$. The standard four-probe technique is 
used for the measurement.    

To ensure uniformity of current, we lithographically 
put wide gold pads on two opposite sides of the sample, each covering the entire 
edge of the network. The transition temperature is measured with a fixed sample 
resistance which is maintained by a feed-back loop with a Linear Research LR-130 
controller. The temperature is measured by a Stanford SR-850 lock-in amplifier 
with a transformer-coupled home-made resistance bridge. The zero-field transition temperature is 
measured to be around $1.2K$ at half of the normal resistance.

In fig.1(a) we show the experimental result of the measured 
phase boundary $\Delta T_C(f)$ of the honeycomb lattice with a locked sample resistance 
$R=R_n/100$, where $R_n$ is the normal-state resistance above the transition. 
As a common procedure \cite{12},  $\Delta T_C$ 
is obtained by a subtraction of the measured $T_C(f)$, defined as the temperature
where $R=R_n/100$, from
 a
smooth parabolic (in $f$) background.  This subtraction
compensates for the critical field of the finite-width wires.

The filling ratio $f=\Phi/\Phi _0$ is  
the magnetic flux $\Phi$ in units of the flux quantum $\Phi _0$($\equiv h c/2e$)
per hexagon.  As predicted by the mean field 
calculation,\cite{11,13} cusps in the $\Delta T_C(f)$
curves are observed at $f=1/3, 2/5, 1/2, 3/5,$ and $2/3$. 

In fig.1(b) we plot the phase boundary of the \kag lattice
for $f$ in the range $[0, 1/8]$, which is measured for the 
same sample resistance ratio of $1/100$. 
Notice here $f$ is the flux per
elementary triangle, instead of per hexagon, and a 
magnetic field of $f=1/8$ corresponds 
to one flux quantum per unit cell 
(a unit cell consists of one hexagon and two triangles).

As we can see from fig.1(a)and (b), there is a one-to-one correspondence of cusps between 
the two phase boundaries. For all the clearly visible cusps in the phase boundary of
honeycomb lattice, i.e.
$f=1/3, 2/5,1/2,3/5,2/3$, we also observe cusps at the corresponding positions in the phase 
boundary of \kag lattice at $f=1/24,1/20,1/16,3/40,1/12$. 
To make the comparison precise, 
if $p/q$ is the value of $f$ for
a cusp in honeycomb phase boundary, then we can 
observe a cusp at $p/8q$ in the \kag
phase boundary.

Since the phase boundary
of the \kag lattice in [0,1] is symmetric about $f=1/2$, 
we will limit our discussion to 
$f \in [0,1/2]$.

The correspondence of the cusps in the phase boundaries also exist between $f \in [0,1]$ of the honeycomb 
lattice and $f \in [1/8,2/8]$ and $[2/8,3/8]$ of the \kag lattice. This correspondence breaks
down for $f \in [3/8,1/2]$, and is discussed below.

The shape and position of the cusps in $\Delta T_C(f)$ curves can be simply understood in 
the following way. An ordered state is usually obtained
by locking fluxoids into a regular commensurate pattern to reduce the energy of the system. 
Although there are a finite number of degenerate states associated with this pattern, which are just 
translations or rotations of the original pattern, they are separated by large energy barriers.      

The addition (or subtraction) of a small number of (extra) flux quanta 
$\delta n$ can be viewed as adding vortices (or antivortices) to the 
locked ground state pattern. 
However, each extra (or absent) flux quantum produces a 
vortex (antivortex) that has a nonzero
energy due to the resulting supercurrents.  Thus the energy of 
the system is increased for either sign of $\delta n$, 
{\em i.e.} $E-E_0 \approx D_{\pm} \left |\delta n \right | $, where $D_+$ is 
the energy cost of each extra flux quantum and $D_-$ is the cost of a removed flux quantum.
So, to lowest order in $\delta n$, the total increase of the energy 
is proportional to the density of the added (subtracted) 
flux quanta, and this in turn leads to the lowering of $T_c$. 
In the end, ordered states correspond 
to local maxima in $T_c$, which appear as downward cusps in the plot of $\Delta T_C$. 

To examine the similarity between honeycomb and \kag more closely, 
let us examine $f=1/3$ (for honeycomb, $f=1/24$ for \kag)
and study the ground-state fluxoid configuration, as illustrated in fig.2. Fig.2(a) is 
the fluxoid pattern for honeycomb at $f=1/3$, and fig.2(b) is the corresponding pattern 
for \kag at $f=1/24$. The shaded hexagons
in fig.2(a) and (b) represent one fluxoid and the unshaded hexagons and triangles represent
zero fluxoid.   To make the comparison more direct, in fig.2(c) we plot
these two networks on top of each other with proper scale between them. As we can see, the fluxoids
in both networks form the same symmetric pattern. 

We can also show that for other values of $f=p/q$ for 
the honeycomb network, the fluxoid configurations for
$f=p/8q$ in the \kag network are the same as the 
corresponding honeycomb configuration.

The similarity in shapes of the phase boundaries between fig.1(a) and fig.1(b) suggests a more 
significant relationship between the two lattices. As shown in fig.2(c), the lattice formed 
by all the hexagons in the
\kag lattice is the same as the honeycomb lattice. Therefore, if, in the \kag lattice, 
all the small triangles are left empty, and fluxoids only stay in the hexagons, we would expect
strong similarities of phase boundaries between the two networks.  Since the area of a hexagon in the
\kag lattice is six times larger than that of a triangle, it costs more energy to add one
fluxoid to a triangle than to a hexagon when $f \in [0,1/8]$ (if we neglect the effects of network).  

The fact that the correspondence with the honeycomb phase boundary exists on the 
\kag for $f \in [0,1/8]$, $[1/8,2/8]$, and [2/8, 3/8],
but not [3/8,1/2], suggests that fluxoids do not go into the triangles in the \kag lattice until
$f=3/8$, regardless of $f$ being rational or irrational 
(for $f>5/8$ we therefore expect 
the triangles to be completely filled). 
Therefore for $f \in [0,3/8]$, the phase boundary is completely determined by the 
configuration of fluxoids in the hexagons which are relatively arranged as in the honeycomb lattice.

A calculation based on quantum interference with multiple-loop 
Aharonov-Bohm Feynman Path-Integral approach is carried out by Lin and Nori \cite{13}. 
Their results, which
are directly connected to the underlying topology, predict exactly the same similarity as seen in 
our experimental data.

Another interesting comparison between the honeycomb 
and \kag superconducting wire networks 
is at $f=1/2$ for both lattices.  As we know, 
both systems have a large degeneracy of ground states at 
half-filling~\cite{15,rz}.  While the phase boundary shows 
a normal cusp at $f=1/2$ for honeycomb, calculation~\cite{11,13,yi}
gives a reverse cusp (a {\it minimum} of $T_c(f)$) for \kag. 
In our experiment, this reverse cusp is observed in niobium \kag
samples,~\cite{yi}
but is smoothed out for our aluminum samples in the
mean field regime~\cite{14}. 

For the \kag lattice, there are many possible ways to accommodate 
$4 \Phi_0$ of total fluxoids
in a unit cell.  In Fig.3(a), we show one ground-state configuration that has
$3 \Phi_0$ per hexagon, $\Phi_0$ in the up-pointing triangles, and no fluxoids in the
down-pointing triangles.  Notice that all the currents in Fig.3(a), as well as in all
other $f=1/2$ ground states are equal in magnitude. Under this condition, we can 
move the fluxoids in the six triangles around one hexagon, as we do from Fig.3(a) to 3(b), 
without changing the configuration elsewhere
and without increasing the energy.  Such local rearrangements can be done anywhere in the
network, demonstrating that the ground-state entropy is proportional to the area of the
network.  If this \kag network is treated in Ginzburg-Landau theory in the limit of
very thin wires, this extensive degeneracy
of the lowest free-energy states persists throughout the superconducting phase.~\cite{park}
In fact, the number of modes that are critical (soft) at $T_c$ in Ginzburg-Landau theory is 
also extensive; a localized soft mode can be made on each hexagon, and in
addition to this ``flat band'' of soft modes, there is one more zero-momentum
soft mode.~\cite{park}  This degeneracy
is an essential ingredient in the reason why this system exhibits
the reverse cusp in its phase boundary at $f=1/2$.~\cite{spectrum}  
It is also worth noting that somewhat similar things
happen at and near $f=1/2$ in the so-called ${\cal T}_3$ lattice, which is
a dual to the \kag lattice~\cite{16,17}.  See also 
Laguna, {\it et al.}~\cite{laguna}, where the same degeneracy is seen for
vortices in a continuum superconductor with a \kag-symmetric pinning
potential.

The honeycomb network at $f=1/2$ has some things in common with the \kag case.
At low temperatures there are an infinite number of ground state current configurations,
as pointed out by Shih and Stroud~\cite{15} for the corresponding Josephson-junction array.
One such pattern is shown in Fig. 4.  In this pattern the fluxoid arrangement in each row
has only two possibilities, but these possibilities may be chosen independently in each row,
resulting in a ground-state entropy proportional to the linear size of the system.  However,
unlike the degeneracy in the \kag network, this degeneracy does not persist throughout the
superconducting phase when the system is treated using Ginzburg-Landau theory.  What happens is
that near enough to $T_c$, when the coherence length is of order the 
lattice spacing or larger, a new,
lower-free-energy pattern enters that has an added nonuniformity in the {\it magnitude} of
the superconducting order parameter.  This pattern is shown in Fig. 5.  The bold arrows indicate
currents flowing along wire segments where the magnitude of the order parameter is relatively
large, while the lighter arrows are currents flowing along wire segments where the magnitude
is smaller.  The magnitudes of the {\it currents} 
are actually identical, so where the order parameter
magnitude is large, the (gauge-invariant) phase gradient is small, and {\it vice versa}.  This
is the pattern that the $f=1/2$ honeycomb network orders into at $T_c$.  It does not have
infinite degeneracy; it just has the finite number of degenerate states obtained from Fig. 5 by discrete
translations and rotations (12 in all).  Thus this is a ``locked'' commensurate state and the
cusp in the phase boundary is of the usual sign, as is seen both in our
experimental results and in Lin and Nori's numerical results.~\cite{13}  
At some temperature below $T_c$, while the
system is crossing from the Ginzburg-Landau regime 
to the London regime (where the order
parameter magnitude nonuniformities are strongly suppressed) 
the system must show a
superconducting-to-superconducting phase transition between the two types of
current patterns shown in Figs. 4 and 5.  We have not yet seriously investigated this
latter phase transition.

In conclusion, we have studied the superconducting-normal phase boundaries of 
honeycomb and \kag lattices. Their shapes are in good agreement with the mean field theory. The position
of the cusps in the phase boundary shows a correspondence between the two types of
networks, which demonstrates the effect of topology of the structure on the phase boundaries. The 
comparison of the two phase boundaries at $f=1/2$ reveals the different natures of their degeneracy.   

We thank Yeong-Lieh Lin and Franco Nori for many discussions and for sending us theoretical
results.   We also thank Kyungwha Park and Roderich Moessner for helpful
discussions.  The work at Princeton University was supported by NSF Grants No. DMR 98-09483
and 98-02468.

\begin{figure}
\caption{Experimentally measured $\Delta T_c(f)$ as functions of $f$.  (a) is for the
superconducting honeycomb network for $f$ in the range between $0$ and $1$.
(b) is for the superconducting \kag network in the range $[0,1/8]$.}
\label{fig1}

\end{figure}

\begin{figure}

\caption{Configurations of fluxoids.  The plaquettes occupied by
fluxoids are shown shaded.  (a) is for the superconducting
honeycomb network with $f=1/3$. (b) is for
the superconducting \kag network with $f=1/24$. (c) shows the two
networks superposed, showing how the fluxoid patterns are identical.}
\label{fig2}
\end{figure}

\begin{figure}

\caption{Configurations of fluxoids and superconducting currents at 
$f=1/2$ for 
two ground states of the \kag network.  The fluxoid arrangement in the
triangles around the central hexagon has been rotated by 60$^o$ between
(a) and (b).}

\label{fig3}

\end{figure}

\begin{figure}

\caption{A ground-state configuration of fluxoids and superconducting currents at $f=1/2$ for 
the honeycomb network at low temperatures, in the London regime.}

\label{fig4}

\end{figure}

\onecolumn
\newpage
\begin{center}

\psfig{file=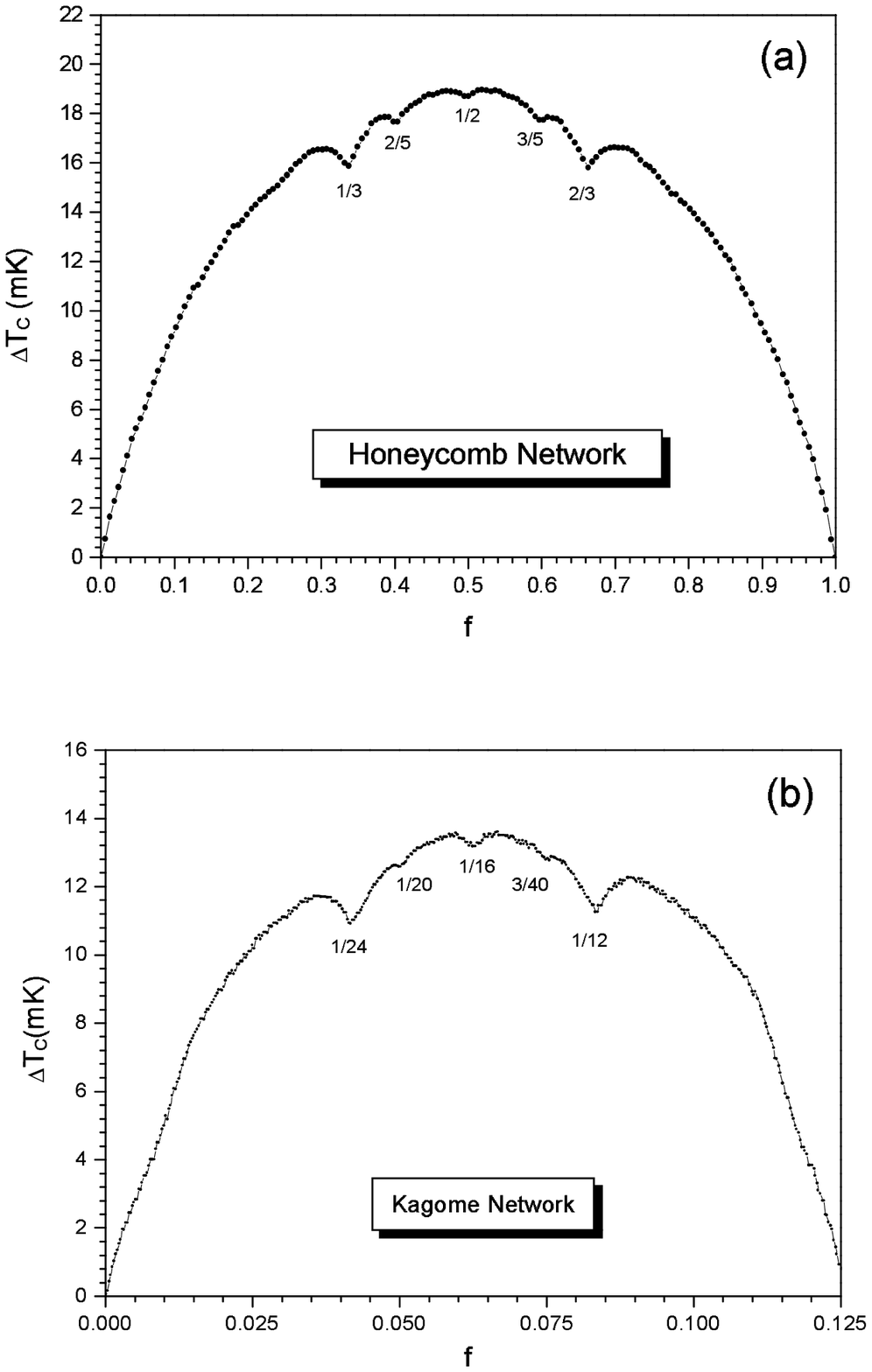,width=5in}\\
\textbf{Fig.1}

\newpage
\psfig{file=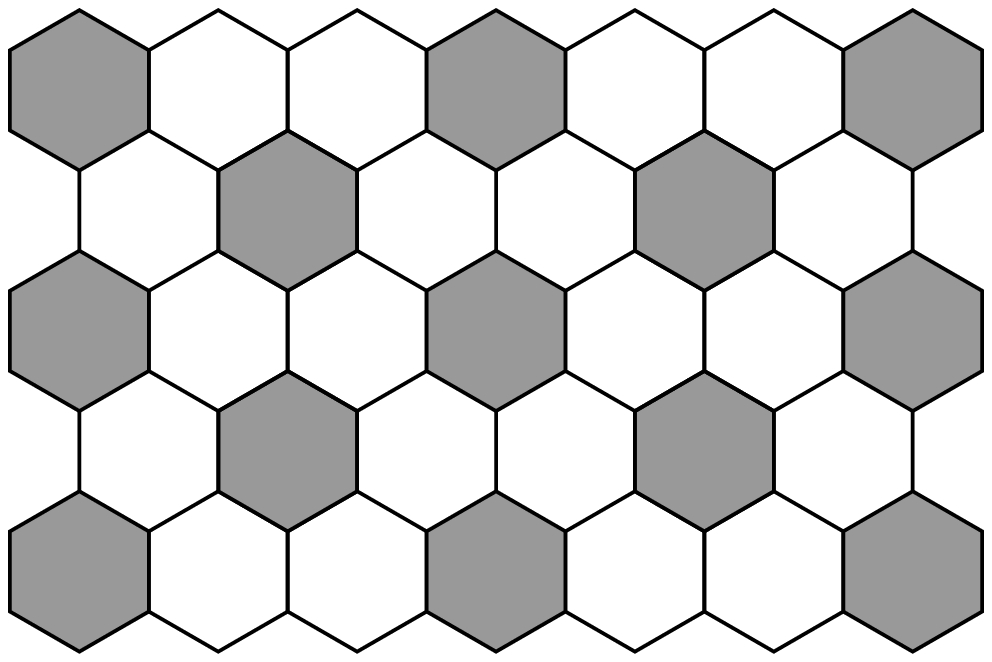,width=3.5in}\\
(a)\\
\vspace{1cm}

\psfig{file=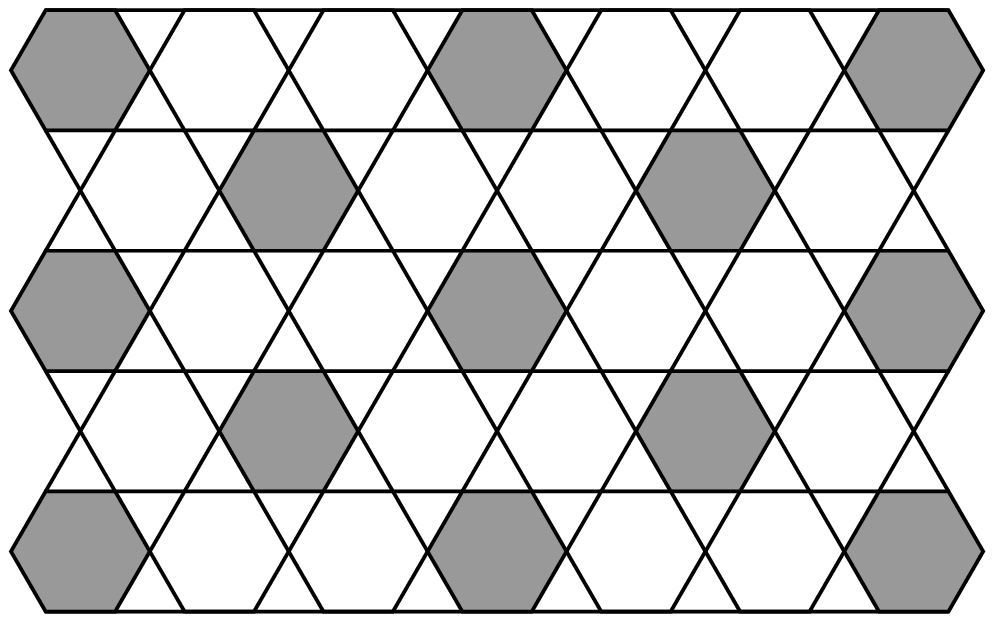,width=3.5in}\\
(b)\\
\vspace{1cm}

\psfig{file=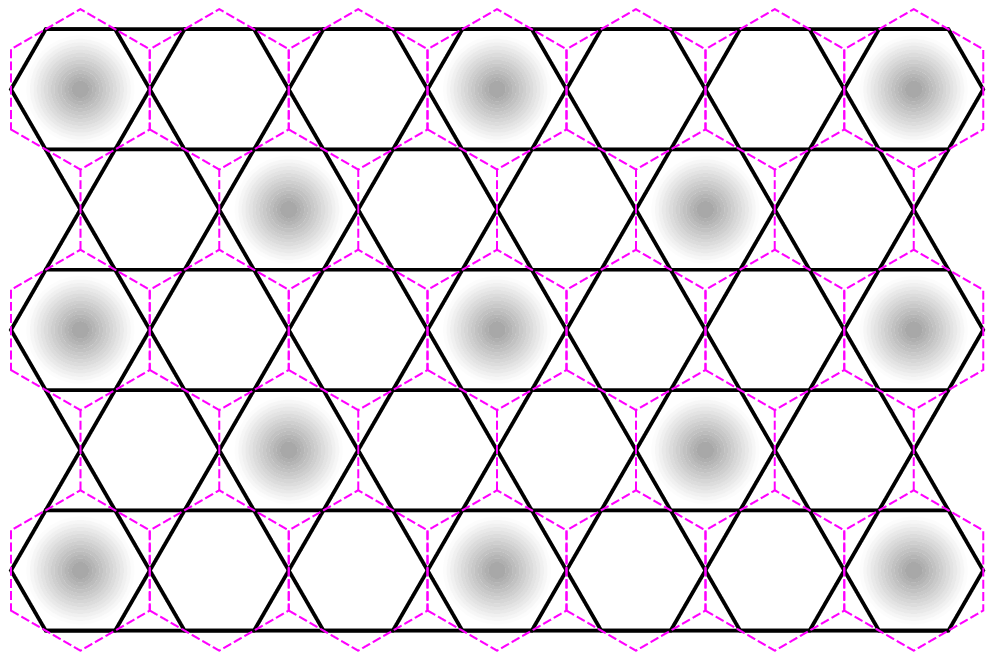,width=3.5in}\\
(c)\\
\vspace{1.5cm}
\textbf{Fig.2}

\newpage
\psfig{file=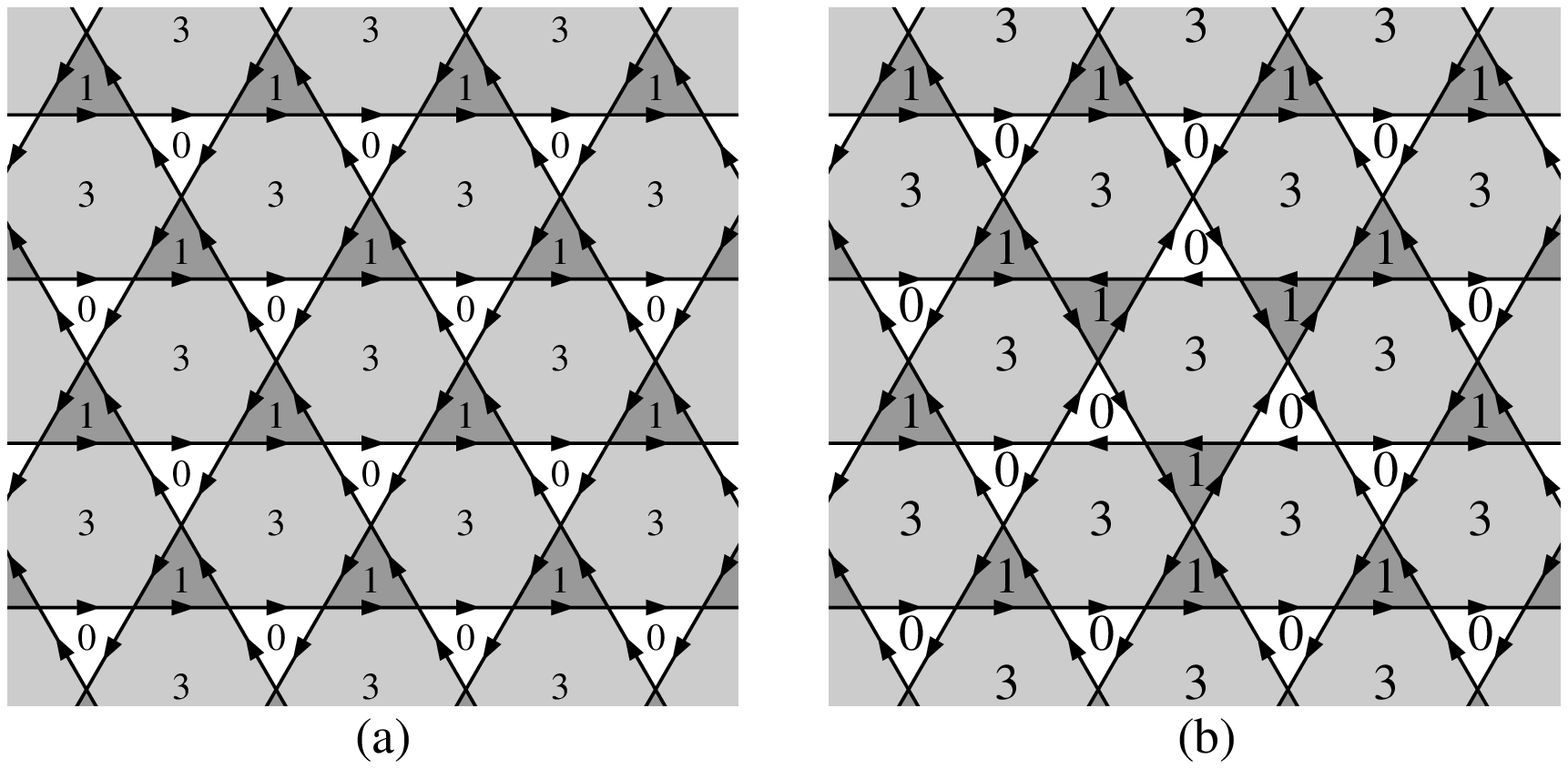,width=4in}\\
\vspace{1cm}
\textbf{Fig.3}

\vspace{2cm}

\psfig{file=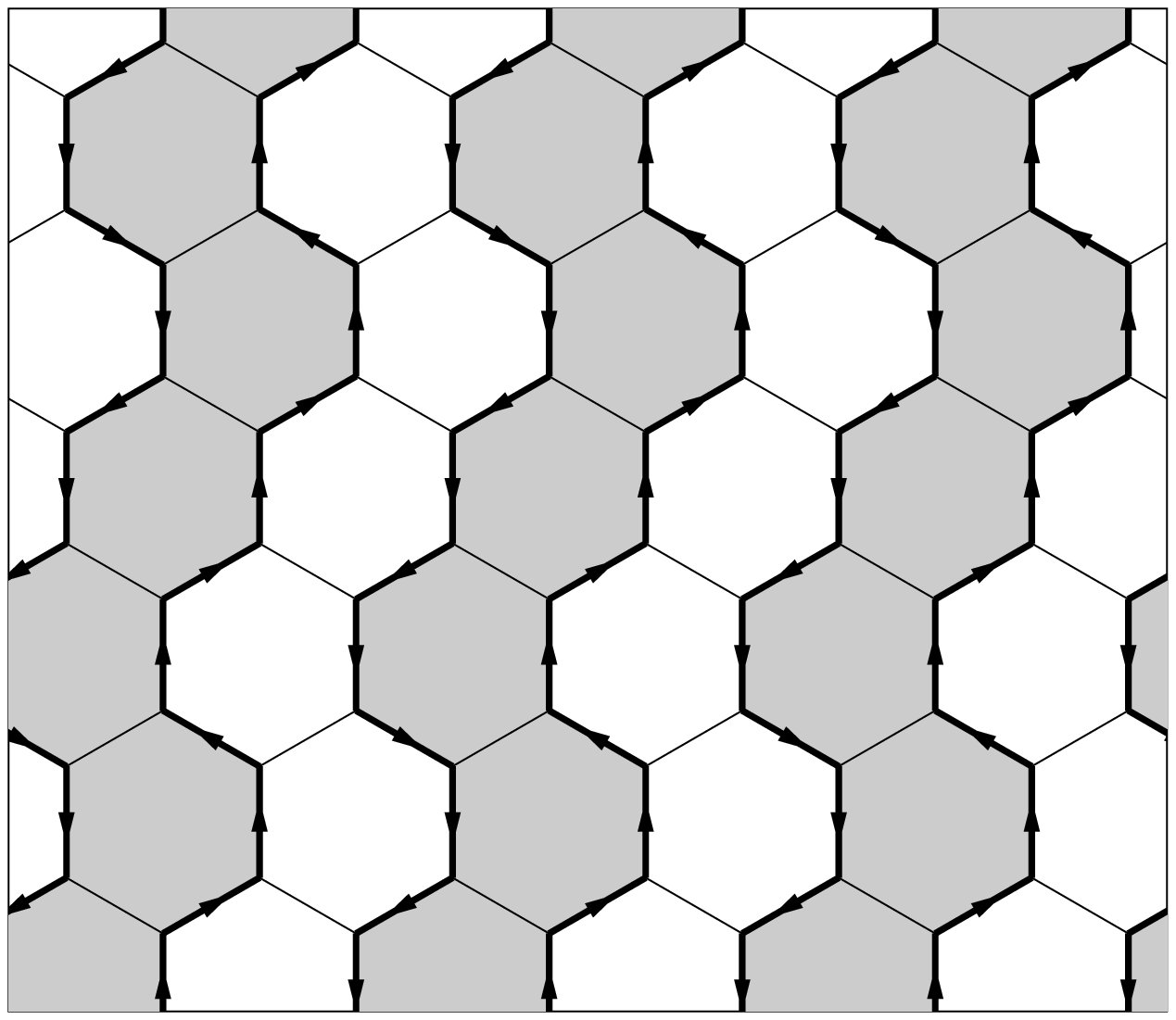,width=4in}\\
\vspace{1.5cm}
\textbf{Fig.4}

\newpage

\begin{figure}
\psfig{file=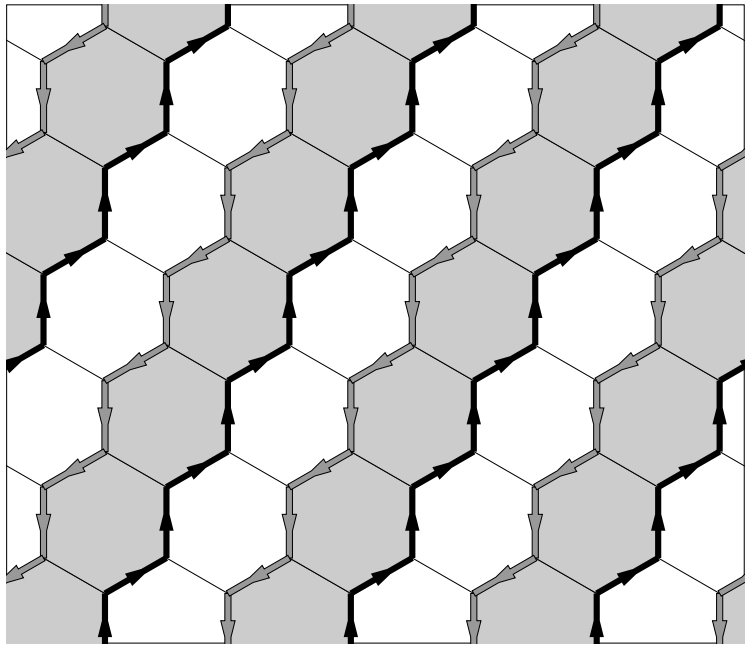,width=4in}\\
\vspace{1.5cm}

\caption{A minimum free energy configuration of fluxoids and superconducting currents at $f=1/2$ for 
the honeycomb network at temperatures near $T_c$, in the Ginzburg-Landau regime.  The magnitude
of the superconducting order parameter is larger on the wire segments with currents indicated
by bold arrows, while it is smaller on the segments where the arrows are shown lighter.}

\label{fig5}

\end{figure}

\end{center}


\end{document}